\documentclass[prb,twocolumn,superscriptaddress,showpacs,floatfix,amsfonts]{revtex4}
\usepackage{graphicx,graphics,color,epsfig}
\usepackage{bm}
\usepackage{amsmath}
\usepackage{amssymb}
\usepackage{epstopdf}
\begin{document}


\preprint{}
\title{Impurity Quantum Phase Transition in a Current-Carrying $d$-Wave Superconductor}
\author{Hua Chen}
\affiliation{Zhejiang Institute of Modern Physics, Zhejiang
University, Hangzhou 310027, China}
\author{Yezheng Wu}
\affiliation{Zhejiang Institute of Modern Physics, Zhejiang
University, Hangzhou 310027, China}
\author{Shuxiang Yang}
\affiliation{Department of Physics and Astronomy, Louisiana State
University, Baton Rouge, Louisiana 70803, USA}
\author{Jianhui Dai}
\affiliation{Condensed Matter Group, Department of Physics, Hangzhou
Normal University, Hangzhou 310036, China}
\affiliation{Zhejiang
Institute of Modern Physics, Zhejiang University, Hangzhou 310027,
China}
\author{Jian-Xin Zhu}
\affiliation{Theoretical Division, Los Alamos National Laboratory,
Los Alamos, New Mexico 87545, USA}
\date{\today}

\begin{abstract}
We study an Anderson impurity embedded in a $d$-wave superconductor
carrying a supercurrent. The low-energy impurity behavior is
investigated by using the numerical renormalization group method
developed for arbitrary electronic bath spectra. The results
explicitly show that the local impurity state is completely screened
upon the non-zero current intensity. The impurity quantum
criticality is in accordance with the well-known Kosterlitz-Thouless
transition.

\end{abstract}
\maketitle


\section{Introduction}

\begin{figure}[t]
\centering \centerline{\includegraphics [width=0.9\columnwidth]
{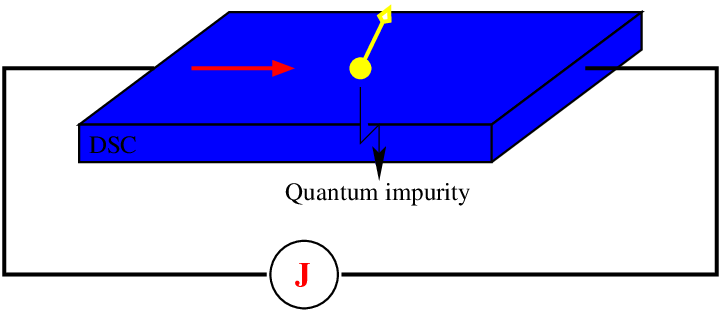}}
 \caption{(Color online) An Anderson impurity embedded in a two-dimensional
$d$-wave superconductor. A
supercurrent $\mathbf{J}$ is fed into the superconductor from a current sourse.
The local electron structure around
the impurity can  be measured by scanning tunneling microscopy.}
 \label{FIG:Intro}
\end{figure}

The behavior of a single magnetic impurity in correlated electron
systems has attracted intensive interest in condensed matter
physics.~\cite{Hewson:Book} While the problem is well studied when
the host is a simple metal,~\cite{Wilson:RMP75} the correlations
among the host electrons may lead to much complicated impurity
behavior deviating from the Fermi liquid
properties.~\cite{Hewson:Book} Specifically, the Kondo screened
state, i.e., an entanglement state between the impurity spin and a
conduction electron, may become unstable at zero temperature upon
the depletion of the density of states (DOS) at the Fermi levels, or
the change of certain nonthermal parameters. Such kind of impurity
quantum phase transition (IQPT) may take place in metallic systems
with a pseudo gap.~\cite{Withoff:PRL90} The response of a known
impurity behavior upon the change of effective couplings could
provide important information of the host itself, and thus can be
used to probe the ground state and low-energy physics of the host
electrons.~\cite{Balatsky:RMP06} On the other hand, within a known
host, the different response of a quantum impurity with internal
dynamical degrees of freedom versus a static impurity, to an
external control parameter, may shed insight on the role played by a
doped impurity in a correlated electron
medium,~\cite{APolkovnikov:01,JXZhu:00} which is a relatively rarely
explored area. In this Article, we study the property of an Anderson
impurity embedded in a $d$-wave superconductor. Our analysis, based
on the numerical renormalization group (NRG) method, unambiguously
shows that an IQPT of the Kosterlitz-Thouless type can be induced in
a $d$-wave superconductor carrying non-zero persist currents. This
means that the Kondo temperature increases with the current $q$
exponentially, $T_K\propto e^{-\alpha /q}$, with $\alpha$ being a
parameter linearly depend on the Coulomb interaction of the impurity
orbital. The obtained results can be tested by scanning tunneling
microscopy (STM), which has been used to measure the local
electronic structure around a doped impurity in
conventional~\cite{AYazdani:97} and unconventional
superconductors~\cite{SPan:00} in the absence of a current flow.

The outline of this paper is as follows. In Sec.~\ref{sec:model}, we
introduce a model system to describe an Anderson impurity in a
current carrying $d$-wave superconductor. We present the formulation
for calculating the quasiparticle spectrum of the superconducting
bath in the presence of the supercurrent. The NRG approach is  then
introduced to solve the quantum impurity problem. In
Sec.~\ref{sec:result}, we present numerical results and identify the
nature of the IQPT. Finally, concluding remarks are given in
Sec.~\ref{sec:summary}.

\section{Model and Method}
\label{sec:model}

It is now well accepted that high-temperature cuprate
superconductors exhibit a $d$-wave pairing
symmetry.~\cite{Xiang:Book} These materials have a two-dimensional
layered structure. In their thin film form, a supercurrent can be
injected by a current source, as shown schematically  in
Fig.~\ref{FIG:Intro}. We notice that a persistent current can also
be generated by piercing a magnetic flux through the axial of a
mesoscopic hollow superconducting cylinder, due to the Aharonov-Bohm
effect.~\cite{JXZhu:94}
The flux driven current is negligibly small
in the thermodynamic limit.  As such, the setup proposed here is
more suitable for the study of quantum impurity problem
in an unconventional fermionic bath, and should be experimentally
accessible.

We model the problem by the following Hamiltonian
\begin{eqnarray}
\mathcal{H}=\mathcal{H}_{\mathrm{BCS}}+\mathcal{H}_{\mathrm{imp}}+
\mathcal{H}_{\mathrm{hybrid}},
\label{EQ:dimp}
\end{eqnarray}
where $\mathcal{H}_{\text{BCS}} = \sum_{\mathbf{k},\sigma}
\xi_{\mathbf{k}+\mathbf{q}} c_{\mathbf{k}\sigma}^{\dagger}
c_{\mathbf{k}\sigma} + \sum_{\mathbf{k}} [\Delta_{\mathbf{k}}
c_{\mathbf{k}\uparrow}^{\dagger} c_{-\mathbf{k}\downarrow}^{\dagger}
+ \text{h.c.}] $, $\mathcal{H}_{\text{imp}} = \sum_{\sigma}
\epsilon_{d} d_{\sigma}^{\dagger} d_{\sigma} +
Un_{d\uparrow}n_{d\downarrow} $, and $\mathcal{H}_{\text{hybrid}}  =
\frac{1}{\sqrt{N_L}}\sum_{\mathbf{k},\sigma} [V_{\mathbf{k}}
c_{\mathbf{k}\sigma}^{\dagger} d_{\sigma} + \text{h.c.} ]$. Here
$c_{\mathbf{k}\sigma}$ annihilates one conduction electron of
momentum $\mathbf{k}$ and spin projection $\sigma$, while
$d_{\sigma}$ annihilates one localized $d$-electron of spin
projection $\sigma$. In the tight-binding approximation, the
conduction electrons have the normal and $d$-wave superconducting
gap dispersions, $\xi_{\mathbf{k}} = -2t (\cos k_x + \cos k_y )
-4t^{\prime} \cos k_x \cos k_y -\mu$ and $\Delta_{\mathbf{k}} =
(\Delta_0/2)(\cos k_x - \cos k_y)$, respectively. The center-of-mass
momentum, $2\textbf{q}$, of a Cooper pair determines the current
flow.~\cite{PGdeGennes} Note that the BCS superconducting part of
the Hamiltonian is written in an unconventional way. We have
transferred the momentum $2\mathbf{q}$ shift on the Cooper pair
wavefunction onto the momentum $\mathbf{q}$ shift on the
single-particle kinetic energy via a local gauge transformation in
real space. A detailed derivation is given in
Appendix~\ref{sec:lgt}. Other parameters $\epsilon_d$, $U$, and
$V_{\mathbf{k}}$ are the localized level, the on-site Coulomb
interaction, and the impurity coupling, respectively. $N_{L}$ is the
number of lattice sites.

The quasiparticle density of states of the 2D current-carrying $d$-wave superconductor can
be found by diagonalizing $\mathcal{H}_{\mathrm{BCS}}$ via the formula
\begin{equation}
\rho(\omega) = \frac{1}{N_{L}} \sum_{\mathbf{k}} [\vert
u_{\mathbf{k},\mathbf{q}}\vert^{2}
\delta(\omega-E^{+}_{\mathbf{k},\mathbf{q}})   + \vert
v_{\mathbf{k},\mathbf{q}}\vert^{2}
\delta(\omega-E^{-}_{\mathbf{k},\mathbf{q}})]\;,
\end{equation}
where
$E^{\pm}_{\mathbf{k},\mathbf{q}} = Z_{\mathbf{k},\mathbf{q}} \pm
[Q_{\mathbf{k},\mathbf{q}}^{2}
+ \Delta_{\mathbf{k}}^{2}]^{1/2}$, with  $Z_{\mathbf{k},\mathbf{q}} =
(\xi_{\mathbf{k}+\mathbf{q}} - \xi_{\mathbf{k}-\mathbf{q}})/2$ and
 $Q_{\mathbf{k},\mathbf{q}}
=  (\xi_{\mathbf{k}+\mathbf{q}} + \xi_{\mathbf{k}-\mathbf{q}})/2$.
The electron- and hole components of the Bogoliubov-de Gennes
eigenfunction are given by
$\vert
u(v)_{\mathbf{k},\mathbf{q}}\vert^{2} = [1 \pm
Q_{\mathbf{k},\mathbf{q}}/E^{0}_{\mathbf{k},\mathbf{q}}]/2$
with $E^{0}_{\mathbf{k},\mathbf{q}} = [Q_{\mathbf{k},\mathbf{q}}^{2}
+ \Delta_{\mathbf{k}}^{2}]^{1/2}$.

In order to solve the whole problem, we generalize the NRG
method~\cite{Wilson:RMP75,Krishna:80} to study the impurity
properties with an arbitrary form of the DOS.
It is sufficient for our purpose to take into account the coupling
between the impurity spin and the particle-excitations. The latter
is the only necessary ingredient in $d$-wave
superconductors.~\cite{Fritz:05,Vojta:02} We therefore ignore the
anomalous part and just study a modified Anderson model with the
impurity coupled to the electron-like excitation spectrum. As such,
we use the following Hamiltonian to the derivation of the NRG
equations:~\cite{Bulla:97,Bulla:RMP07}
\begin{eqnarray}
\mathcal{H}&=&\mathcal{H}_{\mathrm{imp}}+D\sum_{\sigma}\int^{1}_{-1}d\varepsilon
g(\varepsilon)a^{\dagger}_{\varepsilon\sigma}a_{\varepsilon\sigma}
\nonumber\\
&&+D\sum_{\sigma}\int^{1}_{-1}d\varepsilon
h(\varepsilon)(d^{\dagger}_{\sigma}a_{\varepsilon\sigma}+a^{\dagger}_{\varepsilon\sigma}
d_{\sigma})\;, \label{EQ:cimp}
\end{eqnarray}
where we introduced a one-dimensional energy representation for the
particle-like excitations $a^{\dagger}_{\varepsilon\sigma}$ with the
scaled energy $\varepsilon$ and the band-cutoffs at $\pm D$.
$g(\varepsilon)$ and $h(\varepsilon)$ are the energy dispersion and
hybridization self-consistently defined by $\rho(\varepsilon)$ and $V_{\bf k}$
respectively as in Ref.~\onlinecite{Bulla:RMP07}.

\section{Numerical Results}
\label{sec:result}

\begin{figure}
\includegraphics [width=0.9\columnwidth]{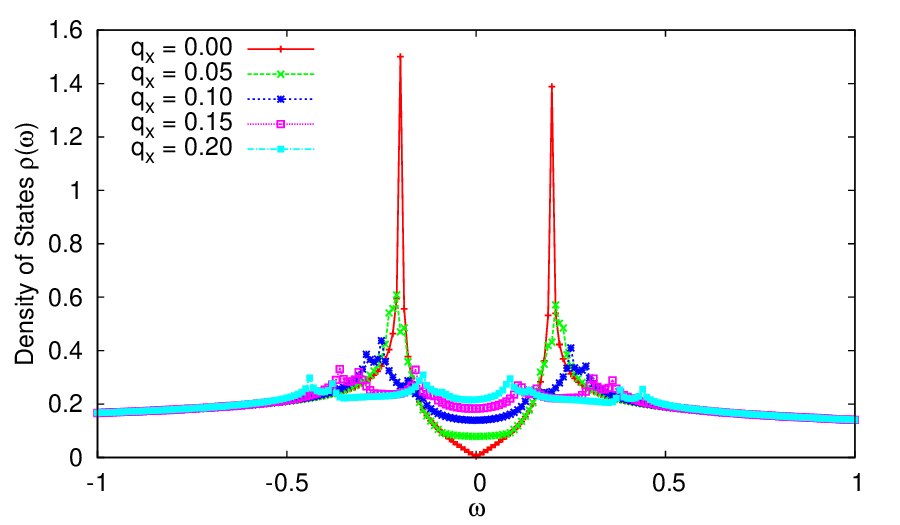}
\caption{(Color online) Density of states as a function of energy
for various values of drift momentum $q_{x}$. } \label{FIG:BulkDos}
\end{figure}

In the numerical calculations, we take $t=1$, $t^{\prime}=-0.2$,
$\mu=-0.78$, $\Delta_{0}=0.2$. The energy is measured in units of
$t=1$ unless specified otherwise. Without loss of generality, we
take $\mathbf{q}=(q_x,q_y)=(q_x,0)$. In Fig.~\ref{FIG:BulkDos}, we
show the DOS as a function of energy for various values of $q_{x}$.
A small intrinsic lifetime broadening $ \Gamma=10^{-3}$, and the
lattice sites of $5000\times 5000$ are chosen. In the absence of the
current, the calculated DOS vanishes linearly as expected, and the
finite size effect on the DOS is negligible as shown in the Appendix~\ref{sec:fs}.
It increases around the Fermi energy as a response to the non-zero
current.

To handle the \emph{arbitrary} $\rho(\omega)$, we follow the generic
scheme of Ref.~\onlinecite{Bulla:RMP07} to discretize the Hamiltonian
given by Eq.~(\ref{EQ:cimp}), and then solve it by the NRG for
different values of $U$ and $\mathbf{q}$ at the symmetric
point $\epsilon_d=-U/2$, with the RG parameter $\Lambda=2$. At
each iteration step we keep about $1500$ states,
depending on the quantities calculated, and the total number of
iterations is $N_{\text{max}}= 100$. Throughout the work, a
value of the band cut-off $D=1$ is chosen.

Distinct from the case of either a wide-band normal metal with a
constant DOS, where the impurity spin being in a strong coupling
(SC) limit, or a conventional $s$-wave superconductor with a
hard-gap everywhere on the Fermi surface, where the impurity spin
being in the local moment (LM) limit, the existence of nodal
zero-energy quasiparticles in a $d$-wave
superconductor~\cite{Balatsky:RMP06,Xiang:Book} has a non-trivial
implication to the fate of quantum impurity states. Earlier studies
by taking the DOS with a soft-gap, $\rho(\omega)\sim
|\omega|^{r}$~($r=1$ for the $d$-wave superconductor), have shown
the existence of a critical coupling, separating the LM and SC
phases.~\cite{Withoff:PRL90,Vojta:PM06,Lee:05,Chen:95,Ingersent:96,Bulla:97}
The present model is more intriguing, as the low-energy excitations
in a $d$-wave superconductor can readily be proliferated by pumping
in a supercurrent, which should have a significant control of the
impurity state.~\cite{note1} The purpose of the present work is to
demonstrate that a Kosterlitz-Thouless-like IQPT can be realized by
tuning the (super-)current.

\begin{figure}
\includegraphics [width=0.9\columnwidth] {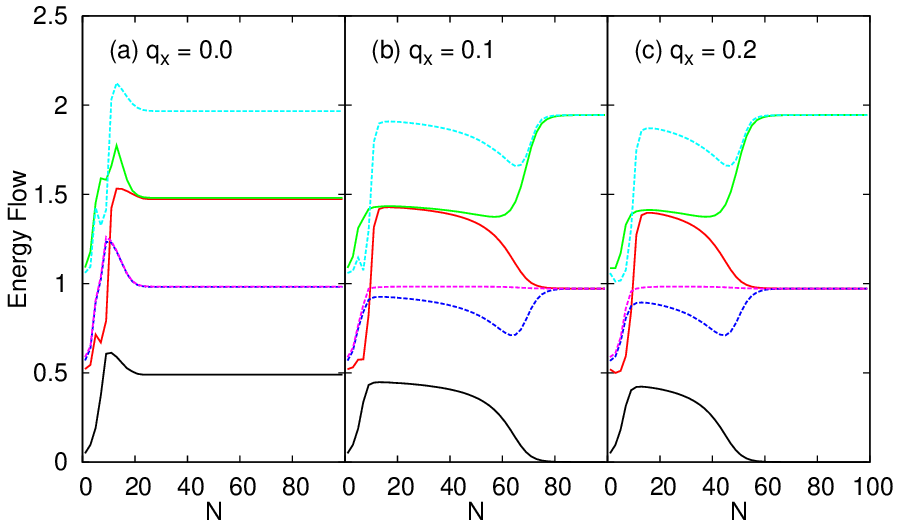}
\caption{(Color online) The NRG energy flows for the low-energy
levels for Coulomb $U=0.1$ with various drift momentum $q_x$. Solid
lines: $(Q,S)=(1,0)$, dashed lines: $(Q,S)=(0,1/2)$.}
\label{FIG:QEF}
\end{figure}

In the absence of the current, the impurity in the present symmetric
Anderson model should be in the LM state for non-zero repulsive $U$
due to the marginal nature ($r=1$) of the $d$-wave superconduting
host. Here we calculate the energy flows as well as other physical
quantities for fixed $U=0.1$ and $V=0.05$, with $q_y=0$ and various
$q_x$. As shown in Fig.~\ref{FIG:QEF}, the NRG flows start and
remain close to the free orbital (FO) regime at the high temperature
or high energy. When lowing the temperature, two types of fixed
points are identified. For the case with $q_x=0$, the many-particle
levels rapidly crossover to the LM fixed point from the FO regime.
For the case with finite $q_x\neq 0$, the SC fixed point develops at
sufficiently low temperatures. The evolution of energy flows from
the LM lineshape to the SC one is similar to the soft-gap Anderson
impurity model, implying that the impurity state is driven into the
SC state by the current.

\begin{figure}
\includegraphics [width=0.9\columnwidth]
{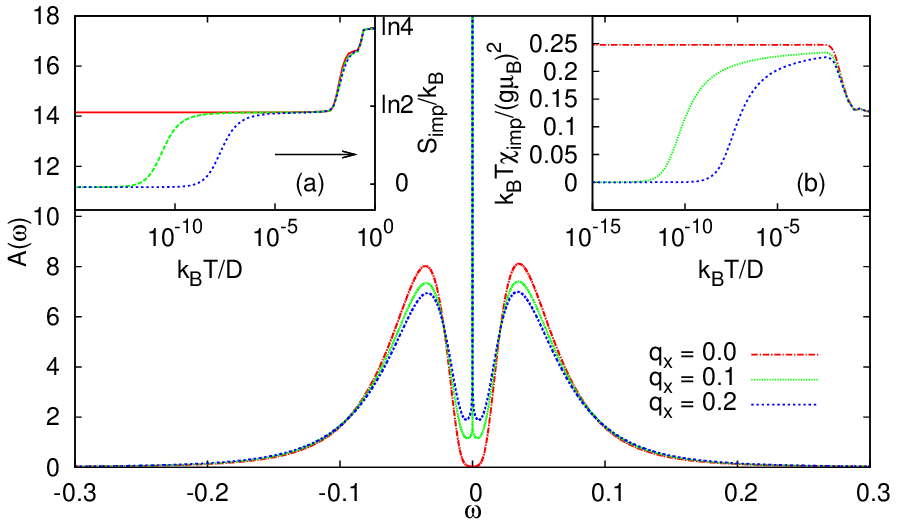} \caption{(Color online) The impurity spectral function at
$T=0$ (main panel) and  the temperature dependence of the effective
impurity moment (right inset) and entropy (left) for Coulomb $U=0.1$
with various $q_x$.} \label{FIG:QSP}
\end{figure}

The IQPT is clearly manifested in some other physical quantities as
illustrated in Fig.~\ref{FIG:QSP}. The impurity spectral function
develops a central peak at zero energy upon increasing $q_x$. This
feature is similar to the soft-gap Anderson model where the impurity
spectral function diverges at the zero energy for the SC and quantum
critical phases.~\cite{Bulla:03} Remarkably, the sum rule is within
98\% accuracy in our case. The IQPT is also indicated in the
temperature dependence of the effective impurity moment and entropy,
plotted in the inset of Fig.~\ref{FIG:QSP}. It shows that the
effective moment and entropy deviate from the free moment values
($0.25{\mu}_B$ and $\ln 2$ respectively) for $q_x=0$ and approaches
to zero for $q_x = 0.1$ and $0.2$. The later two cases indicate a SC
phase where the impurity spin is completely screened at the low
energy scale. We also calculated the finite temperature spectral
function and found that in the SC phase the Kondo peak broadens with
increasing temperatures and ultimately disappears at sufficiently
higher temperatures due to thermal fluctuations.

\begin{figure}
\includegraphics [width=0.9\columnwidth]
{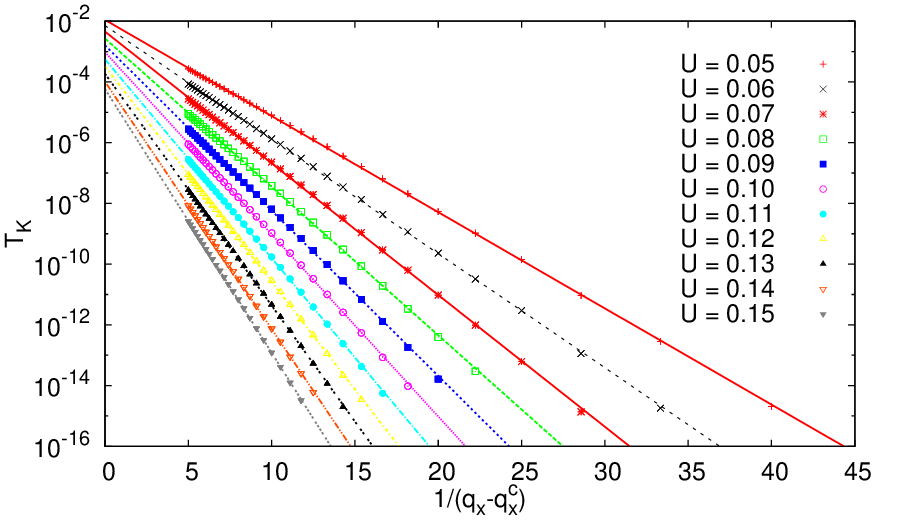} \caption{(Color online) The Kondo temperature as a
function of $1/(q_{x}-q_{x}^c)$ with various Coulomb interaction
$U$. The fitting shows that the critical point $q_{x}^c \to 0$.}
\label{FIG:scaling}
\end{figure}

\begin{figure}
\includegraphics [width=0.9\columnwidth]
{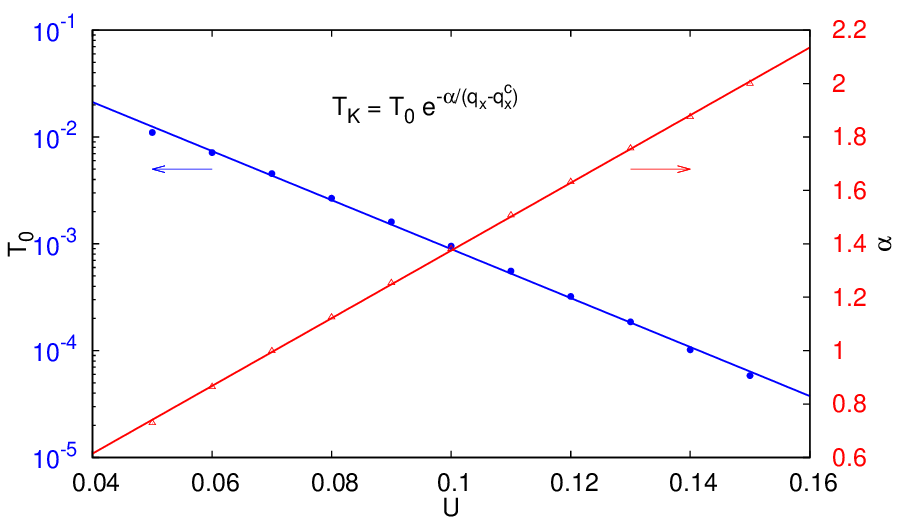} \caption{(Color online) The fitting of Kondo
temperature $T_K=T_0 e^{-\alpha/(q_x-q_x^c)}$, where
$q_{x}^c\rightarrow 0 $, $T_0$ and $\alpha$ are functions of Coulomb
interaction $U$.} \label{FIG:fitting}
\end{figure}

The finite size effect, as detailed in Appendix~\ref{sec:fs}, prevents
us from directly determining the precise location of the critical point
$q_{x}^{c}$, below which the impurity is always in the localized state.
To overcome this difficulty, we perform a scaling analysis on the
critical behavior at the zero temperature limit based on extensive
calculations. We define an energy scale $T_K$, the Kondo
temperature, around which the NRG flows crossover from the LM fixed
point to SC fixed point. Specifically, for each value of
$q_x>q_x^c$, the first excited many-particle energy level drops by
$50{\%}$ at $T_K$~\cite{Bulla:03}. $T_K$ then follows a power-law or
exponential behavior from the criticality, depending on whether it
is a conventional continuous phase transition or a
Kosterlitz-Thouless transition~\cite{Bulla:03}. As shown in
Fig.~\ref{FIG:scaling},  the Kosterlitz-Thouless type nature of the
transition is clearly exhibited. The Kondo temperature $T_K$ follows
a exponential decay with the distance from the criticality, $\ln
T_K=\ln T_0-\alpha/(q_x-q_{x}^c)$, where $T_0$ and $\alpha$ are
functions of $U$, independent on the current $q_x$. Also inferred
from this fitting is that  $q_x^c<5.0\times 10^{-3}$, much smaller
than the scale of $q_x$ ($\sim 0.1$) used in our calculation. Hence
the true location of $q_{x}^c$ is vanishingly small as expected,
implying $q_{x}^{c}\to 0$. The scaling analysis is free from the
finite size effect since the residual DOS vanishes in the
thermodynamic limit. The result therefore shows that
$T_K=T_{0}e^{-\alpha/q_x}$, where, as fitted in
Fig.~\ref{FIG:fitting} for the small U-regime, $\alpha$ increases
linearly with $U$, and $T_0$ decreases exponentially with increasing
$U$.

\section{Concluding Remarks}
\label{sec:summary}

The results reported here explicitly show that the local impurity
state is completely screened upon the non-zero (super-)current
intensity while the impurity quantum criticality is in accordance
with the well-known Kosterlitz-Thouless transition. Our findings
have several implications: (i) In a $d$-wave superconductor,
quasiparticle resonance can be induced around a doped static
impurity in the strong scattering limit. Upon the flowing of a
supercurrent, the resonance peak is suppressed in amplitude and
broadened in width.~\cite{DZhang:05} In contrast, for the quantum
impurity as discussed here, a Kondo resonance emerges when it is
driven into the SC phase by the supercurrent.  This distinction of
the response to the supercurrent can help us to decipher whether a
doped atom plays the role of a static or quantum impurity in a
$d$-wave superconductor. (ii) In high-$T_c$ cuprates like
Bi$_2$Sr$_2$CaCu$_2$O$_{8+\delta}$, the nanoscale inhomogeneity with
the existence of small- and large-gap
domains~\cite{SPan:01,McElroy:Science05} has been ubiquitously
observed by the STM. When a magnetic impurity is doped into the
system, it is in the LM phase in the absence of a supercurrent. In
this case, the impurity plays the role of a weak potential
scatter,~\cite{Kircan:07} and the resonance as due to the
quasiparticle scattering is located far away from the Fermi energy.
This resonance should be observed in both types of domains. When a
supercurrent flows in the sample, the impurities in the small-gap
domains should be first driven into the SC phase and a Kondo
resonance will emerge very close to the Fermi energy while  those in
the large-gap domains are still in the LM with no Kondo resonance.
These interesting phenomena, which can be observed by STM, should
serve as a direct test of our prediction.

We thank R. Bulla, X. Dai and T. Xiang for the technical helps and
useful discussions. H.C., Y. W., and J.D. were supported by the
NSF-China, the NSF of Zhejiang Province, and the 973 Project of the
MOST-China. J.-X.Z. was supported by the National Nuclear Security
Administration of the U.S. Department of Energy at LANL under
Contract No. DE-AC52-06NA25396 and the U.S. DOE Office of Basic
Energy Sciences. Part of numerical simulations were performed on a
computer cluster at the Center for Integrated Nanotechnologies, a
U.S. DOE Office of Basic Energy Sciences user facility.

\appendix

\section{Local gauge transformation}
\label{sec:lgt} A general BCS Hamiltonian can be written as
\begin{eqnarray}
\mathcal{H}_{\text{BCS}} =
&-&\sum_{ij,\sigma}t_{ij}\tilde{c}^\dagger_{i\sigma}\tilde{c}_{j\sigma}
-\mu\sum_{i\sigma}\tilde{c}^\dagger_{i\sigma}\tilde{c}_{i\sigma} \nonumber \\
&+&\sum_{ij}\left[\tilde{\Delta}_{ij}
\tilde{c}^\dagger_{i\uparrow}\tilde{c}^\dagger_{j\downarrow}+\text{h.c.}\right].
\label{eq:bcs}
\end{eqnarray}
In the presence of a supercurrent as carried by paired electrons,
the superconducting pair potential has the form
\begin{eqnarray}\tilde{\Delta}_{ij}=\Delta_{ij}\exp[i\mathbf{q}\cdot \left(
\mathbf{r}_i+\mathbf{r}_j\right)], \label{eq:order}
\end{eqnarray}
where $\mathbf{q}$ determines the center-of-mass motion of Cooper
pairs\cite{PGdeGennes}.

Now we perform the local gauge transformation
\begin{eqnarray}
c_{i\sigma}=\tilde{c}_{i\sigma}\exp[-i\mathbf{q}\cdot\mathbf{r}_i],
\label{gauge}
\end{eqnarray}
the BCS Hamiltonian is given by
\begin{eqnarray}
\mathcal{H}_{\text{BCS}} =
&-&\sum_{ij,\sigma}t_{ij}c^\dagger_{i\sigma}c_{j\sigma}e^{-i\mathbf{q}\cdot\left(\mathbf{r}_i-\mathbf{r}_j\right)}
-\mu\sum_{i\sigma}c^\dagger_{i\sigma}c_{i\sigma} \nonumber \\
&+&\sum_{ij}\left[\Delta_{ij}
c^\dagger_{i\uparrow}c^\dagger_{j\downarrow}+\text{h.c.}\right].
\end{eqnarray}
We then introduce the Fourier transform to cast the BCS Hamiltonian
into the momentum representation,
\begin{eqnarray}
c_{i\sigma}=\frac{1}{\sqrt{N_L}}\sum_{\mathbf k}c_{{\mathbf
k}\sigma}\exp[i\mathbf{k}\cdot\mathbf{r}_i],
\end{eqnarray}
where $N_L$ is the number of lattice sites. A straightforward
algebra yields
\begin{eqnarray}
\mathcal{H}_{\text{BCS}} = \sum_{\mathbf{k},\sigma}
\xi_{\mathbf{k}+\mathbf{q}} c_{\mathbf{k}\sigma}^{\dagger}
c_{\mathbf{k}\sigma} + \sum_{\mathbf{k}} [\Delta_{\mathbf{k}}
c_{\mathbf{k}\uparrow}^{\dagger} c_{-\mathbf{k}\downarrow}^{\dagger}
+ \text{h.c.}].
\end{eqnarray}
This is equivalent to Eq.(1) of the main text in the absence of the
magnetic impurity. In the tight-binding approximation, the
conduction electrons have the normal and $d$-wave superconducting
gap dispersions, $\xi_{\mathbf{k}} = -2t (\cos k_x + \cos k_y )
-4t^{\prime} \cos k_x \cos k_y -\mu$ and $\Delta_{\mathbf{k}} =
(\Delta_0/2)(\cos k_x - \cos k_y)$, respectively. Thus we
demonstrated that the single particle energy picks up a momentum
$\mathbf{q}$-shift from the order parameter. It reflects a
fundamental fact that the superconducting Cooper is formed by two
electrons in the presence of an effective pairing interaction. The
quasiparticle spectrum
\begin{equation}
E_{\mathbf{k},\mathbf{q}}^\pm=\frac{\xi_{\mathbf{k}+\mathbf{q}}-\xi_{\mathbf{k}-\mathbf{q}}}{2}\pm\sqrt{\left(\frac{\xi_{\mathbf{k}+\mathbf{q}}+\xi_{\mathbf{k}-\mathbf{q}}}{2}\right)^2+\Delta_\mathbf{k}^2}
\end{equation}
is easily obtained by diagonalizing the BCS Hamiltonian through a
canonical transformation. It fully agrees with the expression given
in the de Gennes's book.~\cite{PGdeGennes} The same quasiparticle
energy dispersion $E^{\pm}_{{\mathbf k},{\mathbf q}}$, together with
$\xi_{{\mathbf k}}$ and $\Delta_{{\mathbf k}}$, are also given in
the main text.

We emphasize that in the presence of the superconducting order
parameter, the current $\mathbf{q}$ introduced in the expression of
the single particle spectrum is the supercurrent, while it is a
normal state current in the absence of the superconducting order
parameter. This natural recovery merely shows the correctness of the
formalism. Therefore, the supercurrent can tune the quasiparticle
spectrum and density of states in a non-trivial way, which in turn
result in significant consequences on the quantum impurity state in
an unconventional superconducting medium as studied in the main text
of this paper.

\section{The finite size analysis}
\label{sec:fs}

The conduction electron DOS of the bulk system is obtained by
numerical calculation for a relatively large but still  finite
lattice system. There is unavoidably a residual DOS in the absence
of tranport current due to the finite size effect. Since the NRG can
resolve exponentially small energies and temperatures, the residual
DOS at the Fermi level for the $q_x = 0$ case remains a challenging
problem in the finite lattice size calculation and should be treated
with caution. The situation is in contrast to the conventional Anderson
impurity model with a known soft-gap where the analytical expression
of the DOS is available.

\begin{figure}
\includegraphics [width=0.9\columnwidth] {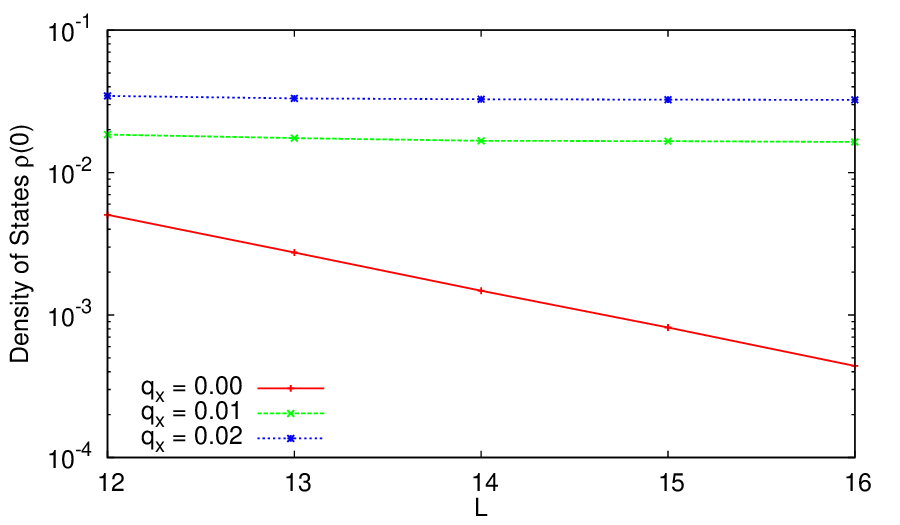}
\caption{(Color online) The DOS at the Fermi level as a function of
$L$ with the system size $N_L = 2^L\times2^L$.} \label{FIG:DOSEF}
\end{figure}

Therefore, we shall clarify how the residual DOS for the $q_x = 0$
case varies with the system size. We first performed a scaling
analysis on the size dependent DOS. To ensure a sufficiently smooth
density of states, we set the broadening parameter $\Gamma
\propto2^{-L}$ in numerical calculations. In the thermodynamic
limit, the intrinsic lifetime parameter $\Gamma \to
0^+$ is recovered. As shown in Fig.~\ref{FIG:DOSEF}, the residual
DOS in the absence of the current becomes very smaller and
approaches zero if we increase the system size up to
$2^{16}\times2^{16}$, while the DOS at the Fermi level for $q_x=0.1$
and $0.2$ is almost size independent. This feature makes our results
based on finite current reliable and robust.

\begin{figure}
\includegraphics [width=0.9\columnwidth] {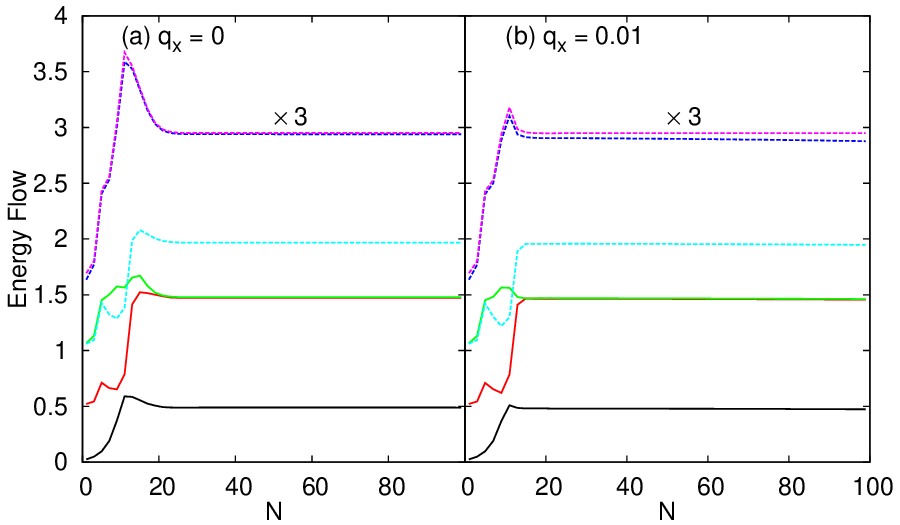}
\caption{(Color online) The NRG energy flows for the low-energy
levels for Coulomb $U=0.05$ with $q_x=0$ and 0.01. Solid lines:
$(Q,S)=(1,0)$, dashed lines: $(Q,S)=(0,1/2)$.} \label{FIG:QEFT}
\end{figure}

We then proceed  to differentiate the low energy behavior,  which may
suffer from the finite size effect for a very small $q_x$. For this
purpose we compare the NRG energy flows for the cases with $q_{x}=0$
and $0.01$, respectively. As shown in Fig.~\ref{FIG:QEFT}, the LM
fixed point is stable up to the maximal iteration $N_{\text{max}}=100$ for
$q_{x}=0$, while it reveals unstable signals (though being tiny) for
very small current $q_x = 0.01$. This provides strong evidence that
the residual DOS at the Fermi level is negligible when the current
is turned on. Hence the scaling behavior of the Kondo temperature
$T_K$ with the finite current investigated in the main text reveals
the correct physics for $q_x\rightarrow 0$.


\begin{thebibliography}{99}
\bibitem{Hewson:Book}A.C. Hewson, \emph{The Kondo Problem to Heavy Fermions},
(Cambridge University Press, Cambridge), 1993.

\bibitem{Wilson:RMP75} K.G. Wilson, Rev. Mod. Phys. {\bf 47},
773 (1975).

\bibitem{Withoff:PRL90} D. Withoff and E. Fradkin, Phys. Rev. Lett. {\bf
64}, 1835 (1990).

\bibitem{Balatsky:RMP06} A.V. Balatsky, I. Vekhter, and J.-X. Zhu,
Rev. Mod. Phys. {\bf 78}, 373 (2006).

\bibitem{APolkovnikov:01} A. Polkovnikov, S. Sachdev, and
M. Vojta, Phys. Rev. Lett. {\bf 86}, 296 (2001).

\bibitem{JXZhu:00} J.-X. Zhu, C. S. Ting, and C.-R. Hu, Phys. Rev. B {\bf 62}, 6027 (2000).

\bibitem{AYazdani:97} A. Yazdani, B. A. Jones, C. P. Lutz, M. F. Crommie, and D. M. Eigler,
Science {\bf 275}, 1767 (1997).

\bibitem{SPan:00} S. H. Pan, E. W. Hudson, K. M. Lang, H. Eisaki, S. Uchida, and J.C. Davis,
Nature (London) {\bf 403}, 746 (2000).

\bibitem{Xiang:Book}T. Xiang, \emph{The D-Wave Superconductors}
(Science Press, Beijing, 2007).

\bibitem{JXZhu:94} J.-X. Zhu and Z. D. Wang, Phys. Rev. B {\bf 50}, 7207 (1994).

\bibitem{PGdeGennes} P. D. de Gennes, {\em Superconductivity of Metals and Alloys}
(W. A. Benjamin, Inc., New York, 1966).

\bibitem{Vojta:PM06}M. Vojta, Phil. Mag.  {\bf 86}, 1807 (2006).

\bibitem{Lee:05}H.-J. Lee, R. Bulla, and M. Vojta, J. Phys.: Conden.
Matter {\bf 17}, 6935 (2005).

\bibitem{Chen:95}Kan Chen and C. Jayaprakash, Phys. Rev. B {\bf 52},
14436 (1995).

\bibitem{Ingersent:96} K. Ingersent, Phys.
Rev. B {\bf 54}, 11936(1996); C. Gonzalez-Buxton and K. Ingersent,
Phys. Rev. B {\bf 57}, 14254 (1998).

\bibitem{Bulla:97}R. Bulla, T. Pruschke, and A.C.
Hewson, J. Phys.: Condens. Matter, {\bf 12},
4899 (1997).

\bibitem{note1}The present model is also distinct from an Anderson impurity
coupled to two s-wave superconducting leads where the IQPT is driven
by varying the magnitude of the hard BCS gap and the phase
difference of the two leads, see in C. Karrasch, A. Oguri, and V.
Meden, Phys. Rev. B {\bf 77}, 024517 (2008).

\bibitem{Krishna:80}H.R. Krishna-murthy, J.W. Wilkins,
and K.G. Wilson, Phys. Rev. B {\bf 21}, 1003 (1980); {\bf 21},
1044 (1980).

\bibitem{Fritz:05} L. Fritz and M. Vojta, Phys. Rev. B
{\bf 72}, 212510 (2005).

\bibitem{Vojta:02} M. Vojta and R. Bulla, Eur. Phys. J. B {\bf 28},
283 (2002).

\bibitem{Bulla:RMP07} R. Bulla, T. Costi, and T. Pruschke, Rev. Mod. Phys.
{\bf 80}, 395 (2008).

\bibitem{Bulla:03}R. Bulla, N.N. Tong, and M. Vojta, Phys. Rev.
Lett.{\bf 91}, 170601 (2003).

\bibitem{DZhang:05} D. Zhang, C. S. Ting, and C.-R. Hu, Phys. Rev. B {\bf 71}, 064521 (2005).

\bibitem{SPan:01} S. H. Pan, J. P. O'Neal, R. L. Badzey, C. Chamon, H. Ding,
J. R. Engelbrecht, Z. Wang,
H. Eisaki, S. Uchida, A. K. Gupta, K.-W. Ng, E. W. Hudson, K. M. Lang, and J. C. Davis,
Nature (London) {\bf 413}, 282 (2001).

\bibitem{McElroy:Science05} K. McElroy, Jinho Lee, J. A. Slezak,
D.-H. Lee, H. Eisaki, S. Uchida,
and J. C. Davis, Science {\bf 309},
1048 (2005).

\bibitem{Kircan:07}M. Kircan, Phys. Rev. B {\bf 77}, 214508 (2008).
\end{thebibliography}
\end{document}